\documentclass{aastex}
\usepackage{amsbsy,amsmath}
\usepackage{multirow}
\newcommand{\bs}[1]{\boldsymbol{#1}}

\newcommand{\be}{\begin{equation}}
\newcommand{\ee}{\end{equation}}
\newcommand{\bear}{\begin{eqnarray}}
\newcommand{\eear}{\end{eqnarray}}

\slugcomment{Manuscript was prepared with AASTeX}

\title{Quantitative Genetics Meets Integral Projection Models: Unification of Widely Used Methods from Ecology and Evolution}

\author{Tim Coulson}
\affil{Department of Zoology, University of Oxford, South Parks Road, Oxford, OX1 3PS} \email{timothy.coulson@zoo.ox.ac.uk}

\author{Floriane Plard}
\affil{Department of Biology, Stanford University, Stanford, CA 94305-5020, USA} \email{Floriane.Plard@stanford.edu}

\author{Susanne Schindler}
\affil{Department of Zoology, University of Oxford, South Parks Road, Oxford, OX1 3PS}
\email{susanne.schindler@zoo.ox.ac.uk}

\author{Arpat Ozgul}
\affil{Institute of Evolutionary Biology and Environmental Studies, Winterthurerstrasse 190, CH-8057 Zurich}
\email{arpat.ozgul@ieu.uzh.ch}

\author{Jean-Michel Gaillard}
\affil{UMR 5558 Biometrie et Biologie Evolutive, Batiment G. Mendel, Universite Claude Bernard Lyon 1, 43 Boulevard du 11 Novembre 1918, 69622 Villeurbanne Cedex, France}
\email{Jean-Michel.Gaillard@univ-lyon1.fr}

\begin{document}
  
\section*{Summary}
\noindent 1) Micro-evolutionary predictions are complicated by ecological feedbacks like density dependence, while ecological predictions can be complicated by evolutionary change. A widely used approach in micro-evolution, quantitative genetics, struggles to incorporate ecological processes into predictive models, while structured population modelling, a tool widely used in ecology, rarely incorporates evolution explicitly.

\noindent 2) In this paper we develop a flexible, general framework that links quantitative genetics and structured population models. We use the quantitative genetic approach to write down the phenotype as an additive map. We then construct integral projection models for each component of the phenotype. The dynamics of the distribution of the phenotype are generated by combining distributions of each of its components.  Population projection models can be formulated on per generation or on shorter time steps.
 
\noindent 3) We introduce the framework before developing example models with parameters chosen to exhibit specific dynamics. These models reveal (i) how evolution of a phenotype can cause populations to move from one dynamical regime to another (e.g. from stationarity to cycles), (ii) how additive genetic variances and covariances (the $\mathbf{G}$ matrix) are expected to evolve over multiple generations, (iii) how changing heritability with age can maintain additive genetic variation in the face of selection and (iii) life history, population dynamics, phenotypic characters and parameters in ecological models will change as adaptation occurs.

\noindent 4) Our approach unifies population ecology and evolutionary biology providing a framework allowing a very wide range of questions to be addressed.  The next step is to apply the approach to a variety of laboratory and field systems.  Once this is done we will have a much deeper understanding of eco-evolutionary dynamics and feedbacks.

Keywords: Additive genetic variance, heritability, integral projection model, eco-evolution, breeders equation

\section*{Introduction}

Evolutionary biologists have a good theoretical understanding of how selection on phenotypes can alter genotype and allele frequencies when the genotype-phenotype map is simple \citep{lande1979quantitative,lande1982quantitative,slatkin1979frequency}.  However, that map can often be complex, particularly when the phenotype is under the control of multiple genes or environmental factors influence phenotypic expression \citep{lynch1998genetics}. The way selection generates evolutionary change can become even more complex when phenotypes change with age due to a mixture of different genes being expressed at different ages and the environmental contribution to the phenotype also varying with age \citep{wilson2005ontogenetic}.  In this paper we develop a framework to study genetic and phenotypic change when the map between genotype and phenotype can be complicated by factors now routinely reported by empiricists.

Research in population genetics has revealed how directional and stabilizing selection erode genetic variation, how disruptive selection, frequency dependence, spatial heterogeneity and heterozygote advantage maintain genetic variation within populations, how drift can generate evolutionary change, and how age structure and temporal environmental variation can slow the loss of genetic variation \citep{crow1970introduction,ellner1994role,slatkin1979frequency}.  The majority of theory demonstrating these processes has used models of a single locus, often with two alleles \citep{charlesworth1994evolution}. In age-structured populations, selection at different ages generates age-specific allele frequencies and genetic correlations across ages. Under some circumstances these age-specific processes can act to maintain genetic variation, although usually they do not \citep{ellner1994role}.  This body of work means that evolutionary biologists now have a deep understanding of how selection can change allele frequencies.

Selection does not operate on genotypes.  Instead it operates on the phenotype, and changes in the phenotype distribution often results in changes to the frequencies of genotypes that determine the phenotype \citep{falconer1960introduction}.  When the map between genotype and phenotype is sufficiently simple, evolutionary change can be inferred by examining changes in the distribution of phenotypes \citep{lande1979quantitative}.  This logic underpins theory behind quantitative genetics -- the study of phenotypic evolution when the genetic architecture of the phenotypic trait is assumed to consist of a large number of genes of small additive effect \citep{falconer1960introduction}.  The equation used to predict evolutionary change in the phenotype within this framework is the breeders equation.  The equation comes in various forms, although all assume that selection on the phenotype directly, and simply, translates into genetic change \citep{lande1979quantitative,charlesworth1993natural,lande1982quantitative}.  Genetic change is assumed to be a simple proportion of change in the phenotypic mean.

The breeders equation estimates change in the mean of the phenotype distribution from one generation to the next.  This is what it was derived to do: to predict evolution in artificial settings if animal and plant breeders were to choose whom to breed from \citep{lynch1998genetics}.  The model is static, not dynamic, in that it should only be used to predict one generation ahead.  The reason for this is that selection alters parameter values in the model, so the heritability (the ratio of the additive genetic variance to the phenotypic variance \citep{falconer1960introduction}) will evolve with time -- something the model does not capture.  Another limitation of the original model is it does not incorporate age structure.  This limitation means the approach cannot be used to gain insight into life-history evolution -- something central to many evolutionary endeavours.  
\cite{lande1982quantitative} developed an age-structured, dynamic version of the breeders equation to address this limitation.  However, in doing this he had to assume that selection is weak and that heritability does not vary with age or time.  The model also assumes that selection on the phenotype proportionally translates to evolutionary change within each age class.  \cite{barfield2011evolution} developed an equivalent approach to Lande for stage-structured populations.

Application of quantitative genetic approaches has worked well in many artificial settings.   One reason for this is that the environment is usually benign, and all individuals experience very similar environments.  Second, artificial selection is usually targeted very precisely at one, or a few, desirable traits.  In nature selection is more diffuse, being targeted at the entire life history \citep{childs2011predicting}.  In addition, the environment is often far from benign, and different individuals can experience very different environments even within the same cohort due, for example, to spatial heterogeneity and variation in maternal effects.  Empirical quantitative geneticists have demonstrated the complexity of environmental contributions to the phenotype by identifying numerous significant environmental drivers when they fit fixed effects into statistical models of phenotypic similarities among related individuals \citep{merila2001explaining}.  For example, age, birth year, population density, climate, maternal condition, dominance, time, spatial variation and the presence of other species can generate environmental variation within the phenotype.  Many of these drivers substantially contribute to the dynamics of the phenotype, both within and between cohorts.  In particular, there is now compelling evidence from several natural systems of heritability varying with age and environment \citep{wilson2005ontogenetic}.  Such complexity is not incorporated into either the breeders equation, or \cite{lande1982quantitative} or \cite{barfield2011evolution}'s extension to it.  A different approach is required, but such an approach needs to build on existing methods.

Understanding the dynamics of phenotypes as well as the dynamics of the underlying genes is important in natural settings.  It is the distribution of phenotypes that determine demography, as it is an individual's phenotype that determines whether it lives or dies, or reproduces or not \citep{coulson2010using}.  However, evolution requires change at the level of the genes.  If we are to understand how anthropogenic actions generate both ecological and evolutionary signatures, it is necessary to understand the dynamics of phenotypes as well as genotypes \citep{coulson2011modeling}.  This is a challenge that is yet to be adequately met in variable environments.

Recently, biologists have developed a powerful framework to study the dynamics of phenotypes, life histories and populations -- integral projection models (IPMs) \citep{coulson2010using,coulson2012integral,easterling2000size,ellner2006integral,steiner2012trading,steiner2014generation}.  However, most IPMs do not incorporate genes (but see \cite{coulson2011modeling}).  This means evolution is usually treated phenomenologically or unrealistic assumptions are made about inheritance.  For example, \cite{coulson2010using} have addressed evolutionary questions with IPMs by asking how evolutionary quantities calculated from models will change when the model is perturbed.  Rees and colleagues \citep{rees2009integral,childs2003evolution} have explored evolutionary end points through recourse to evolutionary game theory.  Both approaches have proved insightful, but some evolutionary biologists have recently raised concerns that these models are not usually evolutionarily explicit \citep{chevin2015evolution}.  

In this paper we combine IPMs with quantitative genetics to develop a general framework for modeling joint phenotypic and evolutionary change.  Specifically, first we show how IPMs can be formulated for the genetic component of the phenotype.  Such IPMs treat inheritance more mechanistically than most IPMs constructed to date, drawing on insights from population and quantitative genetics.   Second, we develop IPMs for the environmental component of the phenotype.  These models incorporate environmental variation in a way that will be familiar to many ecologists.  We then show how the dynamics of the two components can be combined to give the dynamics of the phenotype, the population and the life history.

\section*{Modelling framework}
\subsubsection*{An IPM primer}
Integral projection models are dynamic models that iterate the distribution of a character $N({\cal Z},t)$ forward one time step \citep{ellner2006integral},
\be
	N({\cal Z'},t+1)=\int T({\cal Z}'|{\cal Z},t)w({\cal Z},t)N({\cal Z},t)d{\cal Z} \label{eq:genIPM}.
\ee
\noindent On the right hand side is the initial distribution $N({\cal Z},t)$. This is then operated on by a fitness function $w({\cal Z},t)$ that describes the expected association between a character and fitness.  The product of the initial character distribution and the fitness function gives the distribution of the character post selection $N_s({\cal Z},t)$. This distribution is then operated on by a transmission function $T({\cal Z}'|{\cal Z},t)$ that maps the character distribution post selection to the character distribution in the next time step \citep{coulson2012integral}. 

If the time step used is per generation, the $w({\cal Z},t)$ will describe the association between lifetime fitness and the parental character, and the transmission function will capture the association between parental and offspring characters measured at the same point in the life cycle.  The expectation of $T({\cal Z}'|{\cal Z},t)$ could be estimated from parent-offspring regression or the animal model \citep{lynch1998genetics}.  If the time step used is a shorter interval than the generation, then the fitness function is divided into survival $S({\cal Z},a,t)$ and reproduction components $R({\cal Z},a,t)$ where the association between the character and the demographic rates can vary with age $a$ \citep{coulson2010using}.  Similarly the transmission function $T({\cal Z}'|{\cal Z},t)$ is divided into a function that describes how the trait develops among survivors $G({\cal Z}'|{\cal Z},a,t)$ and how the trait is transmitted between parents and offspring $D({\cal Z}'|{\cal Z},a,t)$.  It is important to appreciate that $D({\cal Z}'|{\cal Z},a,t)$ describes the map between the parental character at time $t$ and the offspring character at time $t+1$ when it recruits to the population in per time step models.  

$T({\cal Z}'|{\cal Z},t)$ is related to $G({\cal Z}'|{\cal Z},a,t)$ and $D({\cal Z}'|{\cal Z},a,t)$ and $w({\cal Z},t)$ to $S({\cal Z},a,t)$ and $R({\cal Z},a,t)$. A consequence of this is that per generation statistics, including selection via lifetime fitness, and the parent-offspring phenotypic covariance, can be calculated from IPMs that operate on timescales shorter than the generation \citep{coulson2010using}.  It is not possible to identify $G({\cal Z}'|{\cal Z},a,t)$, $D({\cal Z}'|{\cal Z},a,t)$, $S({\cal Z},a,t)$ and $R({\cal Z},a,t)$ from per generation IPMs like that in equation \ref{eq:genIPM}. This is why all work to date has been conducted with IPMs that work with a time scale appropriate to the life history of the organism under study (usually a year, but it can be shorter, equation \ref{eq:ageIPM}) \citep{merow2014advancing,coulson2012integral,rees2014building}.

A full age-structured IPM of a multivariate character can be written,
\bear
	N(\bs{{\cal Z}'},1,t+1) &=& \sum_a \int D(\bs{{\cal Z}'}|\bs{{\cal Z}},a,t)R(\bs{{\cal Z}},a,t)N(\bs{{\cal Z}},a,t) d\bs{{\cal Z}} \\
	N(\bs{{\cal Z}'},a+1,t+1) &=& \int G(\bs{{\cal Z}'}|\bs{{\cal Z}},a,t)S(\bs{{\cal Z}},a,t)N(\bs{{\cal Z}},a,t) d\bs{{\cal Z}}, a \ge 1. \label{eq:ageIPM}
\eear
\noindent Here $N(\bs{{\cal Z}},a,t)$ describes the density of individuals with each possible combination of the multivariate character ${\cal Z}_1,{\cal Z}_2,\dots,{\cal Z}_n$ where $n$ is the number of traits under consideration.  The fitness functions $R(\bs{{\cal Z}},a,t)$ and $S(\bs{{\cal Z}},a,t)$ describe how each character influences survival and recruitment at each age, and the transmission functions $G(\bs{{\cal Z}}'|\bs{{\cal Z}},a,t)$ and $D(\bs{{\cal Z}}'|\bs{{\cal Z}},a,t)$ are multivariate probability density functions, describing the probability of transitioning from each multivariate trait value at time $t$ to every possible multivariate trait values at time $t+1$.  IPMs have been extended to the two sex case \citep{schindler2013influence,traill2014demography}.  The methodology we introduce in this paper is extendable to the two-sex case. However, two-sex models contain more functions than those that do not consider males and females separately.  For simplicity, we consider males and females as having identical demographic functions, and we do not work with explicit two sex models.

What form do the functions in the IPM take?  IPMs have been constructed with linear \citep{easterling2000size}, non-linear \citep{ozgul2010coupled}, density-dependent \citep{childs2003evolution}, frequency-dependent \citep{coulson2011modeling} and stochastic functions \citep{rees2009integral}, but here we work with linear and linearised functions. We do this as adding more complexity would require even more parameters than we currently use.  Our approach could be extended to non-linear cases.  Consider a linear function of $\bs{\cal Z}$ at time $t$, $F(\bs{{\cal Z}},t)$, that could take the form,
\be
	F({\bs{\cal Z}},t) = \alpha_0 + \alpha_1 {\cal Z}_1 + \alpha_2 {\cal Z}_2 + \dots \alpha_n {\cal Z}_n + \alpha_{n+1} \int N(\bs{{\cal Z}},t) d (\bs{\cal Z},t) + \dots
\ee
\noindent where the $\alpha$'s are parameters and $\int N(\bs{{\cal Z}},t) d(\bs{{\cal Z}},t)$ is population size.  The function can be extended to add in any other terms the researcher desires.  In the linear case, perhaps where $w(\bs{{\cal Z}},t)$ is being estimated via lifetime reproductive success, it may be sensible for $w(\bs{{\cal Z}},t)$ to be linear $V(\bs{{\cal Z}},t)$.  For survival in an age class, it makes more sense to work on a logit link scale, in which case $S(\bs{{\cal Z}},a,t)=\frac{1}{1+e^{-V(\bs{{\cal Z}},t)}}$ may be a more appropriate form.

IPMs have been used to study the dynamics of populations and life histories structured by phenotypic characters. However, the approach is agnostic to what constitutes the phenotype \citep{smallegange2013towards}.  In contrast, in quantitative genetics, the phenotype is divided into constituent components, including a genetic and non-genetic components \citep{lynch1998genetics}.  Our approach is to use this logic and to write down an IPM for each component of the phenotype.  These IPMs can then be coupled across components.  Because genetic inheritance is mechanistic, and is not influenced by the environment, the transmission functions describing the map between parent and offspring for the genetic component of the phenotype must conform to specific rules.  These rules can be used to capture haploid or diploid inheritance. In addition, the functions can be modified to capture a range of mating systems (see below). The transmission functions for the non-genetic components of the phenotype can be influenced by environmental drivers including population density and weather.  In fact IPMs for the environmental component of the phenotype are typical of IPMs used in ecological settings \citep{coulson2012integral}. Before we introduce the details of our approach, it is useful to consider properties of the transmission functions in further detail, and in particular how the transmission functions can be used to modify the distribution of characters from one age to the next and from one generation to the next. 

\subsubsection*{Some Properties of Transmission Functions}
Our aim in this section is to show haw transmission functions can be used to alter the location and shape of a distribution across ages.  This flexibility allows us to generate any pattern of additive genetic variance across ages we desire, as well as correlations, or lack of correlations, between characters across ages.

Consider a univariate character $\cal Z$. In this example (section), we will ignore selection as it allows us to simplify the notation.  When we apply our approach in all other models in the paper we do include selection.

Assume a Gaussian probability function mapping $N({\cal Z},t)$ to $N({\cal Z}',t+1)$,
\be
	T({\cal Z}'|{\cal Z},t) = \frac{1}{V({\cal Z})\sqrt{2\pi}} e^{-\frac{({\cal Z}'-\mu({\cal Z}))^2}{2V({\cal Z})^2}}.
\ee
\noindent $\mu({\cal Z})$ and $V({\cal Z})$ are functions describing the expected value $\mathbb{E}({\cal Z}',t+1)$ given a specific value of ${\cal Z}$ at time $t$ and $V({\cal Z},t)$ describes the variance around $\mathbb{E}({\cal Z}',t+1)$.  The form and parameterisation of these functions determine similarities and differences between $N({\cal Z},t)$ and $N({\cal Z'},t+1)$.  Assume that $V(\cal Z)$ and $\mu(\cal Z)$ are linear,
\bear
	\mu({\cal Z}) &=& \mu_{0,{\cal Z}} + \mu_{1,{\cal Z}}{\cal Z} \\
	V({\cal Z}) &=& v_{0,{\cal Z}} + v_{1,{\cal Z}}{\cal Z}.
\eear
\noindent The values of $\mu_x$ and $v_x$ can now be used to determine how transmission influences the mean and variance of $N({\cal Z}',t+1)$ given $N({\cal Z},t)$.  To provide an example that should interest quantitative geneticists we split ${\cal Z}$ into a genetic ${\cal G}$ and an environmental ${\cal E}$ component.  We then use transmission functions to increase the mean of the genetic component with age while holding the variance of the genetic component approximately constant.  In contrast, we shrink both the mean and variance of the environmental component of the phenotype with age (Figure 1).  These two changes have the effect of increasing the heritability of the character with age (Figure 2).

In Figure 1 we provide an example where $\mathbb{E}({\cal G})$ increases with age, but the variance remains approximately constant.  The increase in the mean is achieved by placing the function $\mu({\cal G})$ above the $y=x$ line.  The variance changes little as the $\mu_{1,{\cal G}}\approx 1$.  When $\mu_{1,{\cal G}}>1$ then $\sigma({\cal G}',t+1)^2>\sigma({\cal G},t)^2$.  We have also made the transition functions for $\cal G$ deterministic by setting $v_{0,{\cal G}}=V_{1,{\cal G}}=0$: an individual on the $y$th percentile of $N({\cal G},1,t)$ at age 1 will also be on the $y$th percentile of $N({\cal G},a,t+a)$.  In other words, in this example, an individual's breeding value at birth determines its breeding value at later ages although its breeding value does change with age.  However, if independence of breeding values across ages is desired, this too is easily achieved.  A lack of correlation between breeding values at age $a$ and age $a+1$ can be achieved by setting the slope of $\mu_{a,{\cal G}}$ to zero.  Similarly, non-zero correlations can be weakened by increasing the intercept of $V{a,{\cal G}}$. Any desired changes in the means and variances of breeding values across ages can be achieved through the choice of parameters in the development function.

In contrast to the breeding value transition functions described in our example, those for $\cal E$ are probabilistic. This is achieved by making $v_{0,{\cal E}}>0$.  Values of $v_{0,{\cal E}}>0$ add uncertainty in transitions injecting additional variation into $N({\cal E},t)$ as trajectories are no longer deterministic as in our breeding value example.  To counter variance added by $v_{0,{\cal E}}>0$, variance is reduced by setting $\mu_{1,{\cal E}}<1$. $\mathbb{E}(\cal E)$ reduces with each age by keeping the majority of the transition function below the $y=x$ line.  If we were to assume an additive genetic map such that $\cal Z = G + E$ then these functions act to increase the heritability of $\cal Z$ with age.  In addition, $\mathbb{E}(\cal Z)$ increases with age.

This example shows how flexible transmission functions can be, and how they can easily be used to generate specific variance-covariance structures between characters measured at different points during life.  We now outline our approach, showing how this flexibility can be used to combine structured models and quantitative genetics.

\subsection*{IPMs and quantitative genetics}
In this section we work through each of the steps required to construct an IPM of a phenotypic character with a known (or assumed) genotype-phenotype map.

\subsubsection*{Define the genotype-phenotype map}
Start by defining a genotype-phenotype map.  A simple case can be written, 
\be
	{\cal Z}={\cal G}+{\cal E}
\ee
\noindent where ${\cal G}$ is the breeding value of an individual and ${\cal E}$ is the environmental component of that individual's phenotype.  A more complex case could assume the breeding value and environmental components of the phenotype interact 
\be
	{\cal Z}={\cal G}+{\cal E}+\beta_1 {\cal G}{\cal E}.  
\ee
The environmental component of the phenotype could depend upon the environment, perhaps as a function of population density $N(t)$ at time $t$, and some stochastic driver $\epsilon(t)$: 
\be
	{\cal E}=\beta_2N(t)+\beta_3\epsilon(t).
\ee
\noindent The map can be as flexible as desired, and can assume breeding values (as above) or any genetic architecture desired. For example, \cite{coulson2011modeling} have already developed models where the genotype is determined by genotype at a single diallelic locus. Although we only consider two components of the phenotype, it can be divided into as many components as desired.

Next we write a model for the dynamics of $\cal G$ and $\cal E$.  We start by working out the consequences of selection, before focusing on transmission of each component of the phenotype.

\subsubsection*{Estimate how selection modifies the distribution of each component of the phenotype}
\noindent Selection operates on the phenotype $\cal Z$.  Consider, to start, the expected fitness $w({\cal Z},t)$ of an individual with trait $\cal Z$ as 
\be
	w({\cal Z},t)=\alpha_{0,{\cal Z}}+\alpha_{1,{\cal Z}}\cal Z. 
\ee
\noindent The association between the trait and fitness can also be mediated by density or any other aspect of the environment, 
\be
	w({\cal Z},t)=\alpha_{0,{\cal Z}}+\alpha_{1,{\cal Z}}{\cal Z}+\alpha_{2,{\cal Z}}N(t)+\alpha_{3,{\cal Z}}\epsilon(t).  
\ee
\noindent In this case we have assumed the fitness function is linear, but this assumption is relaxed below.  

Next, work out how selection operating on the phenotype impacts each component of the phenotype. When ${\cal Z}={\cal G}+{\cal E}$ then the slope $\alpha_{1,{\cal G}}$ of $w({\cal G},t)$ will be $\alpha_{1,\cal Z}$.  However, if the mean of the environmental component of the phenotype $\mathbb{E}({\cal E})\ne0$ then the intercept of the fitness function for $\cal G$ will not be $\alpha_{0,{\cal Z}}$.  We can find out what it is by writing $\mathbb{E}(w({\cal Z},t))=\alpha_{0,{\cal G}}+\alpha_{1,{\cal G}}\mathbb{E}({\cal G})$ and solving for $\alpha_{0,{\cal G}}$.  The reason we can do this is $\mathbb{E}(w({\cal G},t))$ must equal $\mathbb{E}(w({\cal Z},t))$.  If it did not, then the breeding value distribution $N({\cal G},t+1)$ would differ in size to the phenotype distribution $N({\cal Z},t+1)$.

Similar logic can be used for the more complex cases.  For example, when ${\cal Z= G + E}+\beta_1\cal GE$ then the slope of the fitness function for $\cal G$, $w({\cal G},t)$, will be 
\be
	\alpha_{1,{\cal Z}}+\alpha_{1,{\cal Z}}\beta_1\mathbb{E}({\cal E},t)
\ee
\noindent and the intercept will be
\be
	\alpha_{0,{\cal Z}}+\beta_1\mathbb{E}({\cal E},t).
\ee
\noindent An identical approach is used to construct a fitness function for the environmental component of the phenotype $w({\cal E},t)$. Once functions that describe how selection alters the distribution of each component of the phenotype have been identified, the next step is to construct the functions that describe transmission of each component of the phenotype.  We start with the simple case where we work on a per generation time step.

\subsubsection*{Transmission of components of the phenotype from parents to offspring}
We have already introduced the transmission functions above. However, our focus was on how distributions change with age -- something we will return to in a later example.  In this section we think about transmission from parent to offspring.

In a per-generation IPM (equation \ref{eq:genIPM}) the transmission function describes the map between parental and offspring traits both measured at the same age.  If we were to construct the function for the genotype, $\mu(\cal Z)$ would be a parent-offspring regression, or the intercept and heritability estimate from the animal model.  However, we do not directly estimate the function $\mu(\cal Z)$.  Instead we define one transmission function for $\cal G$, $T({\cal G'}|{\cal G},t)$ and another for $\cal E$, $T({\cal E'}|{\cal E},t)$.  The choice of parameters for $T({\cal G'}|{\cal G},t)$ are not estimated from data, but are instead determined by the inheritance mechanism, the mating system, and any additional assumptions the researcher wishes to impose.  

We will consider the inheritance of breeding values.  In a simple haploid case, breeding values are passed from parent to offspring with perfect fidelity.  This can be achieved with the following parameterisation $\mu({\cal G})={\cal G}$ and $V({\cal G})=0$. Mutation can be added by generating variation around $\mu({\cal G})$ by setting parameters in $V({\cal G})$ such that $V({\cal G})>0$.  

A diploid transmission function assuming random mating requires some assumptions.  It is usually assumed in quantitative genetics that the mean of the offspring breeding value distribution is equal to the mean of the parental breeding value distribution $\mathbb{E}({\cal G'})=\mathbb{E}({\cal G})$ \citep{falconer1960introduction}.  We will also assume that the variances are equal -- i.e. transmission does not lead to a change in the genetic variance $\sigma({\cal G},t)^2$.  This can be achieved with the following parameterisation: $\mu({\cal G})=\frac{\mathbb{E}({\cal G},t)}{2}+0.5{\cal G}$ and $V({\cal G})=0.75\sigma({\cal G},t)^2$.

In Figure 2 we can see how haploid and diploid random mating transmission functions generate identical distributions of $N({\cal G'},t+1)$ when $N({\cal G},t)$ is Gaussian.  Different parameterisations of the diploid transmission functions can capture other mating systems and inheritance processes.  For example, assortative mating will increase the slope of $\mu_{1,{\cal G}}$ to be above 0.5 while dis-assortative mating will decrease it.  The evolutionary consequences of different matings systems can easily be investigated within our framework but are beyond the focus of this paper.  Instead we now turn our attention to parent-offspring transmission functions in age-structured models.

We have already considered transmission functions among survivors that can alter the distribution of $\cal G$ and $\cal E$ from one age to the next (Figure 1). What will the parent-offspring transmission functions $D({\cal G'}|{\cal G},a,t)$ look like?  The logic will be similar to that for the per-generation functions described in the paragraphs above in this section.  A parent with a particular breeding value at age $a$ will not necessarily have had that breeding value at birth.  It is both the birth breeding value, and the expected trajectory of breeding values throughout life, that need to be passed from a parent to its offspring.  There are various ways in which this can be achieved, but probably the most straightforward is to keep track of the multivariate distribution of each individual's age-specific breeding values.  In cases where the development of the breeding value is deterministic, as in Figure 1, the birth breeding value determines a deterministic breeding value trajectory throughout life.  In contrast, if the transmission functions among survivors are probabilistic, the breeding value trajectory throughout life will also be probabilistic: each birth breeding value will have a distribution of possible trajectories, each with a different likelihood.

Once a genotype-phenotype map has been defined, selection on each component of the phenotype has been derived from selection on the phenotype, and transmission functions have been constructed for each component of the phenotype, we have the various pieces of the jigsaw to iterate the IPMs for each component of the phenotype forward from one time step to the next.  But how can we construct what is happening to the phenotype?

\subsubsection*{Dynamics of the phenotype}
We start by calculating moments of each of the components of the phenotype at time $t+1$.  For example, the mean $\mathbb{E}({\cal G}',t+1)$ of $N({\cal G}',t+1)$ can be calculated as,
\be
	\mathbb{E}({\cal G}',t+1) = \frac{\int{\cal G}'N({\cal G}',t+1)d{\cal G}'}{\int N({\cal G}',t+1)d{\cal G}'}.
\ee
\noindent while the variance $\sigma({\cal E}',t+1)^2$ would be calculated as,
\be
	\sigma({\cal E}',t+1)^2 =  \frac{\int{\cal E}'{\cal E}'N({\cal E}',t+1)d{\cal E}'}{\int N({\cal E}',t+1)d{\cal E}'} -\left(\frac{\int{\cal E}'N({\cal E}',t+1)d{\cal E}'}{\int N({\cal E}',t+1)d{\cal E}'}\right) ^2.
\ee
\noindent In multivariate cases genetic and environmental covariances between characters can also easily be calculated.  Once the desired statistics have been calculated from each component they can be combined given the genotype-phenotype map to give desired moments of the phenotype distribution $N({\cal Z}',t+1)$.  For example, to calculate the mean (standard deviation) of $N({\cal Z}',t+1)$ for an additive map ${\cal Z} = {\cal G} + {\cal E}$ simply sum together the two means (standard deviations).  It is, in fact, possible to combine transmission functions across components of the phenotype to generate phenotypic transmission functions, but space precludes us from elaborating on this here.

Having introduced the logic of our framework, we now provide two examples.  The first is a simple bivariate per-generation model based on the same assumptions as the multivariate breeders equation. It allows us to study the dynamics of the $\mathbf{G}$ matrix. The second is more elaborate, involving density-dependent feedbacks and age-structure. In all models, parameter values are selected to give the types of dynamics we wish our models to exhibit.  Parameter values are not based on any particular study system.

\section*{Example models}
\subsubsection*{Model to demonstrate a per generation IPM}
We consider two traits, ${\cal Z}_1$ and ${\cal Z}_2$, and an additive genotype-phenotype map: $\bs{\cal Z = G + E}$. The distributions $N(\bs{{\cal G}},1)$ and $N(\bs{{\cal E}},1)$ are both bivariate normal.  The means and covariance matrices defining these distributions are given in Appendix 1.

Next we assume a linear fitness function without density-dependence or environmental stochasticity,
\be
	w(\bs{{\cal Z}},t) = 0.1+0.15{\cal Z}_1-0.042{\cal Z}_2.
\ee
\noindent This results in the following fitness functions for the additive genetic and environmental components of the phenotype,
\bear
	w(\bs{{\cal G}},t) &=& 0.1+0.15\mathbb{E}({\cal E}_1,t)-0.042\mathbb{E}({\cal E}_2,t) + 0.15{\cal G}_1-0.042{\cal G}_2 \nonumber \\
	w(\bs{{\cal E}},t) &=& 0.1+0.15\mathbb{E}({\cal G}_1,t)-0.042\mathbb{E}({\cal G}_2,t) + 0.15{\cal E}_1-0.042{\cal E}_2
\eear
\noindent Finally we define two bivariate transmission functions -- one for $\bs{\cal G}$ and one for $\bs{\cal E}$.  These both take the form,
\be
	T(x'_1,x'_2|x_1,x_2,t) = \frac{1}{2\pi \sqrt {V(x_1) V(x_2)(1-\rho^2)}} e^{\frac{-\omega}{2(1-\rho^2)}}. \label{eq:gaussian}
\ee
\noindent where
\be
	\omega = \frac{(x'_1-\mu(x_1))}{V(x_1)}+\frac{(x'_2-\mu(x_2))}{V(x_2)} - \frac{2\rho(x'_1-\mu(x_1))(x'_2-\mu(x_2))}{\sqrt{V(x_1)V(x_2)}}
\ee
\noindent with $\rho=\frac{cov(x_1,x_2)}{\sigma(x_1)\sigma(x_2)}$ being the correlation coefficient between $x_1$ and $x_2$.  We use the following parameterisations in our first model (model $1A$) assuming haploid inheritance, $\mu({\cal G}_1) = {\cal G}_1$, $V({\cal G}_1) = 0$, $\mu({\cal G}_2) = {\cal G}_2$, $V({\cal G}_2) = 0$, $\mu({\cal E}_1) = \mathbb{E}({\cal E}_1,t)$, $V({\cal E}_1) = \sigma({\cal E}_1,t)^2$, $\mu({\cal E}_2) = \mathbb{E}({\cal E}_2,t)$ and $V({\cal E}_2) =  \sigma({\cal E}_2,t)^2$.

\noindent This has the effect of keeping the distribution of the environmental component of the phenotype constant from one generation to the next.  Next we extend this model to create a diploid model (model $1B$).  To do this we simply modify the parameterisation of the transmission function for the genetic component of the phenotype $\mu({\cal G}_1) = \frac{\mathbb{E}({\cal G}_1,t)}{2}+0.5{\cal G}_1$, $V({\cal G}_1) = 0.75\sigma({\cal G}_1,t)^2$, $\mu({\cal G}_2) = \frac{\mathbb{E}({\cal G}_2,t)}{2}+0.5{\cal G}_2$ and $V({\cal G}_2) = 0.75\sigma({\cal G}_2,t)^2$.

At each time point we calculate means and variances for each trait, and the covariance between traits, for $N(\bs{{\cal G}},t)$ and $N(\bs{{\cal E}},t)$, along with total population size $n(t) = \int N(\bs{{\cal G}},t) d\bs{{\cal G}}$.  The change in fitness between generation $t$ and $t+1$ can then be calculated as $R0=\frac{\int n(t+1)dn}{\int n(t)dn}$.  We also calculate the means and variances of $N(\mathbb{\bs{\cal Z}},t)$ by summing together the appropriate means and variance-covariance matrices.  For plotting purposes we assume the distribution of the phenotype at a given time point is bivariate normal, constructing it from the means and variance of each trait and the covariance between them.

The means of both phenotypes evolve with time (Figure 4a). Similarly, we see small decreases with time in the additive genetic variances and covariances as selection slowly erodes it (Figure 4b).  Fitness, measured as mean lifetime reproductive success $R0$, increases with time (Figure 4c).  These changes are reflected in simultaneous changes in the distributions of both the phenotype and the additive genetic variance, with both evolving in the same directions (Figure 4d). These results occur because the environmental component of the phenotype remains unchanged from one generation to the next.  Altering the inheritance mechanism from haploid (Model $1A$) to diploid (Model $1B$) has relatively little effect on results.  The mean of each phenotype changes a little more slowly in the diploid model than in the haploid one, while the genetic variances and covariances change a little faster.  However, after 60 generations the differences are hardly noticeable. Given the small differences, and because the haploid model has fewer parameters, for the remainder of the paper we focus on haploid cases.

In both the haploid and diploid model the breeding value and phenotype distribution continue to change at a rate that only slows gradually.  The reason for this is our initial distributions $N({\cal G},1)$ and $N({\cal E},1)$ have tails that tend to zero but never quite get there. This means that there is a very small density of extreme breeding values.  Over time, the densities of these rare breeding values grow as selection increases their frequency.  In order to stop this process, we can make some parts of the breeding value distribution off-limits to evolution \citep{blows2005reassessment}.  We do this by forcing the initial values of the distribution to zero.  This is one way of imposing a genetic constraint in the haploid model -- the breeding value distribution cannot enter this area of evolutionary space.  When we do this, we can see that evolution pushes the distribution of the breeding value up against the constraint in one dimension.  A consequence of this is the evolution of mean fitness slows, and the genetic covariance between the two traits changes from being negative to being positive as the genetic variance in one dimension is eroded towards zero.  This change is also reflected in the distribution of the phenotype (Figure 4L).

During the course of the simulations, parameters in some of the functions can change.  For example, as the mean breeding value at generation $t$ $\mathbb{E}(\bs{[{\cal G}},t)$ evolves in model $1B$ so too do the intercepts of $\mu({\cal G}_1)$ and $\mu({\cal G}_2)$. In our simulation the intercept for trait 1 increased from 3.5 at $t=1$ to 5.56 at $t=60$, while the intercept for trait 2 decreased from 4 to 2.74.  Similarly the intercepts of $V({\cal Z})$ also evolved over the course of the simulation (trait 1: 0.75 to 0.53, trait 2: 0.53 to 0.44). Although the fitness function did not change during the course of the simulation, the intercept of the function describing the strength of selection on the environmental component increased from 0.814 to 1.55. However, because the environmental component of an offspring is not inherited by its offspring, this selection is not realised in direct evolutionary change.  Exactly which parameters will change value, and to what extent, will depend upon details of the simulation.  For example, in the haploid model $1A$ the intercept of $\mu({\cal G})$ will not change during the course of the simulation because offspring inherit their parent’s ${\cal G}$ with perfect fidelity and $\mu({\cal G})={\cal G}$.  Parameters also change in our next age-structured model. Our aim with this model is to show how age-structure and density can interact to generate some interesting ecological and evolutionary dynamics.

\subsection*{An age-structured density-dependent model}
We consider a pre-breeding census, and three age-classes: juveniles (aged one year) which do not breed, prime aged adults aged between two (the earliest age an individual can breed) and seven, and a senescent class aged eight and above.  The genotype-phenotype map at birth is additive with ${\cal Z} = {\cal G} + {\cal E}$.  Selection occurs in each age-class and is density-dependent.  Breeding values develop with age in a near deterministic manner, with a low probability of mutation.  The environmental component of the phenotype shrinks with age as in Figure 1.  The dynamics of the environmental component are also density-dependent.  Technical details of each of the models we describe below are provided in Appendix 1.  Our aim here is simply to showcase the types of ecological and evolutionary patterns models within our framework can capture.

We start each simulation with the same initial conditions -- a cohort of juveniles.  Different parameterisation of the model (Appendix 1) can generate different dynamics. In all models, and not unexpectedly, the mean breeding value increases with time. This is because selection is always directional, with parameters describing the effect of the character on survival and fertility being positive within each age class.  

In our first model (model $2A$) evolution leads to a continuous, but small, change in population size, population structure and mean demographic rates (and hence life history). These demographic patterns occur because selection increases the mean of the breeding value and the phenotypic trait, while decreasing the size of the environmental component of the phenotype.  The additive genetic variance decreases during the course of the simulation while the heritability increases with age. These changes occur as parameter values within the model change (Figure 5a).  These results are satisfying, as they support previous work showing how ecological and evolutionary quantities are necessarily linked \citep{coulson2010using,coulson2011modeling}.  However, they do not reveal any particularly surprising patterns.

Directional selection, as in model $2A$, usually erodes additive genetic variation.  We have already shown (Figure 1) that development functions can be used to increase additive genetic variance within a cohort as it ages.  However, because offspring inherit the parental breeding values that they expressed at birth (and with perfect fidelity in our haploid model), changing breeding values with age do not influence the breeding value the parent passes to its offspring.  Changing breeding with age can impact the strength of selection, and consequently the additive genetic variance in a birth cohort, but this effect appears weak in our models (results not reported).  However, another process can inject additive genetic variance into each new generation.  In the discussion that follows we focus on the additive genetic variance at birth, but similar results can be generated for any age (Appendix 1).

In model $2C$, selection erodes the birth additive genetic variance within a cohort as it ages (Figure S1).  However, selection also increases the difference between means in birth additive genetic variances across ages from one year to the next.  The rate at which selection increases the difference in these means, and consequently increases the between age-class birth additive genetic variance, is greater than the rate at which selection erodes the birth additive genetic variance within a cohort as it ages.  The birth additive genetic variance is a sum of the average variances within each age classes and the variance of the means across age classes. When evolution increases the second of these at a faster rate than at which it erodes the first, then evolution will increase the additive genetic variance at birth.  In order to demonstrate this we simply modified model $1A$ by increasing the strength of selection via survival.  This had the effect of increasing the differences in the mean additive genetic variance across ages within a cohort as it aged.  The effect in our models is small because we chose parameter values close to the switch point at which evolution decreases, and increases, the additive genetic variance as viability selection is altered. 

Our third model (model $2C$) is parameterised to demonstrate how evolution can lead to population dynamics changing from one type of dynamical pattern to another.  We do this by changing a number of parameter values to increase the density-dependence to become over-compensatory.  In addition, we increase the strength of selection compared to model $2A$ (Appendix 1). Selection of this strength is stronger than that generally reported in natural systems.  However, we do this to generate striking dynamical change within the 200 year simulation.  In this model we see that as evolution leads to an increase in the trait mean we see the population dynamics evolve from a two point to a four point cycle (Figure 5(B)).  These changes are accompanied by a shift in the dynamics of the population structure and survival rates. As in model $2A$ we see an increase in the mean of the phenotype and breeding value distribution and an initial increase, and then decrease, in the environmental component of the phenotype.  We see the additive genetic variance increase during the course of the simulation for the same reason as reported in model $2B$.

There are many other ways that our models can be modified to generate a suite of dynamics that we will cover in later papers.  For example, some parameterisations can result in evolution increasing the likelihood of extinction \citep{j2005can}, while others reveal the conditions under which cryptic evolution \citep{merila2001cryptic} will occur.

\section*{Discussion}
In this paper we develop a framework to model the evolutionary dynamics of phenotypic traits in variable environments.  Our approach permits phenotypes to be determined additively, non-additively, linearly and non-linearly by genes (or breeding values) and the environment.  Selection can be linear (model $1$), or non-linear (model $2$), and a function of the environment.  We give an example of a density-dependent environment, but stochasticity can just as easily be added as is routinely done in other integral projection models \citep{coulson2012integral}.  Transmission of genetic and non-genetic components of the phenotype can be flexibly defined, capturing a range of mating systems and inheritance mechanisms.  Phenotype distributions can be iterated forwards on a per generation time step, or on time steps that are shorter than a generation.  The framework we have developed provides a way to link ecological and evolutionary dynamics.  We provide toy models to demonstrate the wide range of ecological and evolutionary dynamics our framework can generate.

Our work substantially extends a body of literature showing how ecological and evolutionary dynamics can be simultaneously investigated across a range of assumptions about the way phenotypes, or life histories, are inherited.  For example, \cite{coulson2010using} showed how ecological and evolutionary dynamics can be jointly explored phenomenologically by asking how quantities of interest to ecologists and evolutionary biologists change when parameters in structured models are altered. \cite{coulson2011modeling} then went on to demonstrate how simple genetic architectures can be incorporated into the IPM framework, linking ecological models and population genetic models.  \cite{childs2011predicting} used IPMs coupled to game theoretic approaches to compete life histories against one another assuming an offspring number-offspring size trade-off.  This approach assumes life history strategies (rather than a phenotypic trait) are clonally inherited.  In the work we present here we take a step to unifying these contrasting approaches.  First, we show how parameters are expected to change as evolution occurs. Second, because the breeding value at birth determines breeding values at later ages, each birth breeding value has its own phenotype, but also its own age-specific expected phenotypes, survival and reproductive rates. As the phenotype evolves, the life history evolves too.  There are still several useful developments to extend the IPM approach further to investigate joint ecological and evolutionary dynamics, but we believe, in spite of recent criticism \citep{chevin2015evolution}, that structured modelling is a powerful tool to unify ecology and evolution, and certainly the most promising one currently available.

One of the most exciting aspects of our work is the way that evolution can be included into models that have primarily been used by ecologists.  Our modelling framework can capture many of the dynamics that have interested population ecologists for decades.  For example, our models can be used to generate cycles and chaos by increasing reproductive rates to a point at which density-dependence becomes overcompensatory \citep{may1974biological}. When models are parameterised to be close to a threshold between two dynamical patterns, it is possible for phenotypic evolution to push the population across this threshold (Model $2C$). This property makes it possible to address questions about how phenotypic character evolution and population dynamics are linked. The link occurs because changes in the distribution of phenotypic characters lead to changes in demographic rates.  If changes in these rates modify the shape of the density-dependence \citep{coulson2008estimating} making it sufficiently non-linear, then populations can move between dynamical regimes.  However, even in models that are not close to dynamical regimes, evolution generates changes in population dynamics -- for example, mean $R0$ (the population growth rate on a per generation time step: model 1) and carrying capacity in density-dependent models (mode 2A). Evolution of phenotypic characters, life histories and evolutionary parameters are clearly intimately linked, even if in some instances changes are relatively small (see also \citep{coulson2010using,coulson2011modeling})

But our insights are not only ecological in flavour.  In another demonstration of the utility of our approach, we identify how changing heritability with age can contribute to the maintenance and generation of additive genetic variance among newborns. Empiricists have previously identified evidence that the additive genetic variance can differ with age, with heritability sometimes being greater at older ages than at younger ones \citep{wilson2005ontogenetic}. However, previously it has not been straightforward to incorporate such structure into models \citep{lande1982quantitative}. We have shown that when this structure is included we find that the way that breeding values change with age can either increase or decrease the rate of evolution of character depending upon the way the environment influences selection, and via which component of fitness (demographic rate) it operates.

What is it about our approach that is novel, and why does it allow ecological and evolutionary dynamics to be linked in a way that could not be done in earlier models?  The approach we outline in this paper also builds on seminal work by \cite{lande1982quantitative}.  Lande showed how quantitative genetic parameters can be linked to demography to predict life history evolution by combining additive genetic variances, selection gradients and sensitivities of the population growth rate to the demographic rate via which selection was acting.  Without this important insight we would not have seen how to use IPMs to link ecological and evolutionary dynamics.  However, our work builds substantially upon Lande's important contributions.  

A key breakthrough that IPMs allowed was the ability to work with entire distributions, rather than just moments of the distribution.  Most previous quantitative genetic approaches focus solely on the dynamics of the mean phenotype (or occasionally the variance) \citep{falconer1960introduction, lynch1998genetics}.  In these evolutionary models, the dynamics of the mean depend upon the variance (and the dynamics of the variance depend upon the dynamics of the skew).   These traditional models consequently require assumptions about moments higher than the one under study in order to allow multi-generational predictions \citep{rice2004evolutionary}.  This is the primary reason why modelling the dynamics of the $\mathbf{G}$ matrix has proven so challenging \citep{arnold2008understanding}.  Modelling the dynamics of the entire distribution avoids this issue of dynamic insufficiency \citep{coulson2010using}.  Tracking the dynamics of entire distributions is not complicated; it is also one of the most appealing aspects of IPMs \citep{ellner2006integral}. The IPM approach also allows the consequence of the assumptions used in classical quantitative genetic models to be tested -- a useful focus of future work.  We suspect that when selection is weak, classic models will probably perform well.  

Lande \citep{lande1982quantitative}, and researchers who followed his lead \citep{barfield2011evolution}, have worked with moments of the joint breeding value and phenotype distributions, but have not explicitly considered the dynamics of the environmental component of the phenotype.  Because ${\cal G}$ is a component of ${\cal Z}$, and because ${\cal G}$ is inherited with mechanistic rules that are not impacted by the environment, while inheritance of the environmental component can be impacted by the environment \citep{falconer1960introduction}, it is at best difficult, and may be impossible, to incorporate the effects of the environment on the dynamics of phenotypic inheritance by focusing on ${\cal G}$ and ${\cal Z}$.  It is instead necessary to focus on the dynamics of $\cal G$ and $\cal E$ and to derive the dynamics of $\cal Z$ from these.

Another key insight our framework allows is how model parameters are expected to evolve with time. The extent to which parameters change will depend upon details of the simulation (see results) and clearly further work is required to understand how evolution is expected to impact parameters within models. However, the results we obtained do provide a further insight.  

If evolution occurs over the time period a population is studied, and is not corrected for in statistical analyses, it can bias results.  Consider, for example, the case of a phenotypic trait evolving from having a mean of two to a mean of four during the course of a study.  Now assume we have a density-dependent fitness function of the form,
\be
	w({\cal Z},t) = \alpha_0+\alpha_1{\cal Z}+\alpha_2N(t).
\ee
\noindent If the fitness function was parameterised at the beginning of the study, but the phenotypic trait was ignored as is often the case in ecological studies, then the fitness function would take the form,
\be
	w({\cal Z},t) = (\alpha_0+2)+\alpha_2N(t).
\ee
\noindent In contrast, at the end of the simulation, the function would take the form,
\be
w({\cal Z},t) = (\alpha_0+4)+\alpha_2N(t).
\ee
\noindent Because the intercept of the phenotype-free fitness function has evolved during the course of the study, if the ecological model
\be
w({\cal Z},t) = \alpha_3+\alpha_2N(t)
\ee
\noindent was fit to data from the entire duration of the study, the intercept would clearly lie somewhere between $\alpha_0+2$ and $\alpha_0+4$ and the estimate of the slope $\alpha_2$ would be biased.  Of course, not all phenotypic traits are measured, and temporal trends of year could be fitted to models as is sometimes done to look for evidence of evolution.  However, in times of environmental change it is impossible to disentangle evolution and environmental change caused by trending unmeasured environmental variables in such a case. 

Exactly the same criticism can be levelled at dynamic models of evolutionary change.  If density, or another environmental variable, changes during the course of a study, but is not included into predictive models, the models will give biased predictions of phenotypic change.  This is a likely cause of the failure of application of the breeders equation in the variable environments of nature, but why it succeeds in the constant environmental settings of the laboratory and the greenhouse \citep{morrissey2010danger}.  Once again, year could be included in models as a continuous fixed effect to correct for such trends, but again it confounds potential evolutionary and environmental change.  Fortunately our approach provides a way to include both environmental variation and evolution.  However, the question arises do we really need all this complicated machinery?
 
Compelling evidence of evolution over a small number of generations in the lab is rare. However, a few cases have been demonstrated, including guppies in Trinidad where evolution from a high predation to a low predation environment takes approximately 20-30 generations \citep{reznick1997evaluation}. Our simulations do reveal evolution of the mean phenotype on such a time scale, but we have deliberately chosen heritabilities that are much higher than those typically reported from the wild.  In those models which also exhibit the greatest evolutionary change, we also impose selection that is stronger than typically observed in the field.  We consequently caution against claims of evolution in a generation or two \citep{coltman2003undesirable} as likely being flawed given our results suggest a few tens of generations are required before compelling evidence of evolution is likely to be detectable (see also \cite{hadfield2010misuse}).

It this hypothesis is correct, then it is probably appropriate to model the population, life history and phenotype dynamics of most systems using IPMs that do not explicitly incorporate evolution. This is because statistically detectable changes in parameter values are likely to take many generations. Phenomenological analysis of IPMs that are not evolutionarily explicit will likely provide robust insight \citep{coulson2010using,ellner2006integral,merow2014advancing,rees2014building}.  However, for those systems where evolution is known to occur, like the Trinidadian guppies, the full machinery of the framework outlined here will be required.  Clearly further work and parameterisation of models within our framework to real systems, and comparison with simpler models, is required before our hypotheses can be tested.

There are several directions future work should address. First, having derived the framework and theoretically shown its potential, we need to show how it can be parameterised for real systems.  In reality, this is relatively straightforward, and be can done using the animal model \citep{lynch1998genetics} and other standard regression methods \citep{crawley2002statistical}.  Second, we need to validate models. The way to do this would be to try to predict simultaneous ecological and evolutionary change as populations adapt to a novel environment. Obviously this requires species that are known to exhibit rapid evolution in the field and the lab.  Finally, there are a number of other interesting questions that can be asked.  In developing the framework we realised we were able to address the following:
\begin{itemize}
\item{What are the consequences of adaptive and non-adaptive phenotypic plasticity on character evolution, life history evolution and population dynamics?}
\item{How will species interactions impact joint ecological and evolutionary dynamics?}
\item{How does asymmetric competition in trait-mediated IPMs \citep{bassar2015sub} generate frequency-dependence and impact ecological and evolutionary dynamics?}
\end{itemize}

There are doubtless other important questions our framework can address. But while these are addressed, or we seek funds and fellows to address them, we hope that both ecologists and evolutionary biologists embrace our framework, as it really does allow the phenotypic, life history and ecological consequences of evolution to be explored in a very flexible manner.  And surely that is a key step in generating a unifying theory of ecology and evolution.  

\section*{Acknowledgements}
TC acknowledges past support from the NERC and the ERC that contributed to the development of this approach.  Thanks too to Shripad Tuljapurkar and Julia Barthold for providing useful comments on a first draft of the manuscript.

\bibliography{root}

\clearpage

\begin{figure}
\centering
\includegraphics[width=9cm]{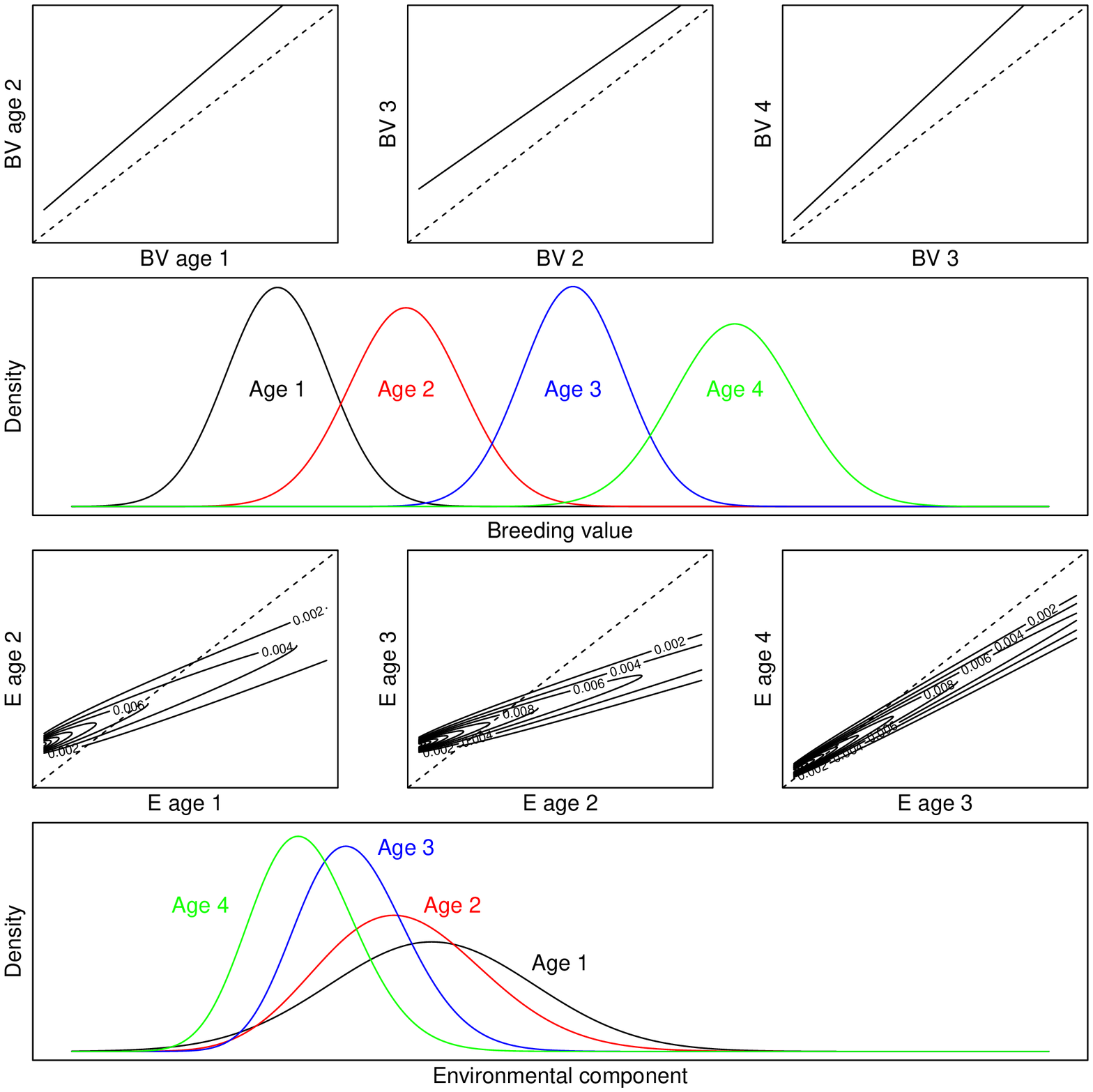}
\caption{\label{figure:fig1} Using transition functions to modify distributions.  In order to increase the mean of a distribution from one time step to the next, the majority of the $\mu({\cal Z})$ function must be above the $y=x$ line. In the panel in the top row we grow the mean of a breeding value distribution across four ages.  The resulting dynamics are reproduced in the panel in the second row. If $\mu_{1,{\cal Z}}>1$ the variance in the distribution will increase from one time step to the next.  In the panels in the third row, we generate functions to reduce the mean of the environment component of a phenotype. These are probability density functions, with parameters in the variance function $V({\cal Z})$  greater than 0. This acts to inject variance into the distribution of the environmental component of the distribution, partially countering the variance that is lost due to $\mu_{1,{\cal Z}}<1$ these transition functions act to increase the heritability of the character with age, and will also generate non-zero genetic and phenotypic covariances across ages.}
\end{figure}

\clearpage

\begin{figure}
\centering
\includegraphics[width=9cm]{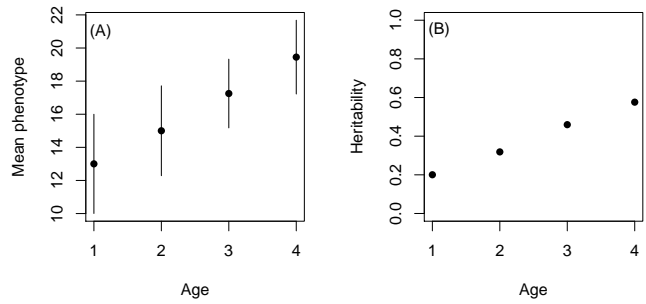}
\caption{\label{fig:fig2} The transmission functions graphed in Figure 1 generate (a) a phenotype that increases in mean value with age and (b) a heritability that increases with age.}
\end{figure}

\clearpage

\begin{figure}
\centering
\includegraphics[width=9cm]{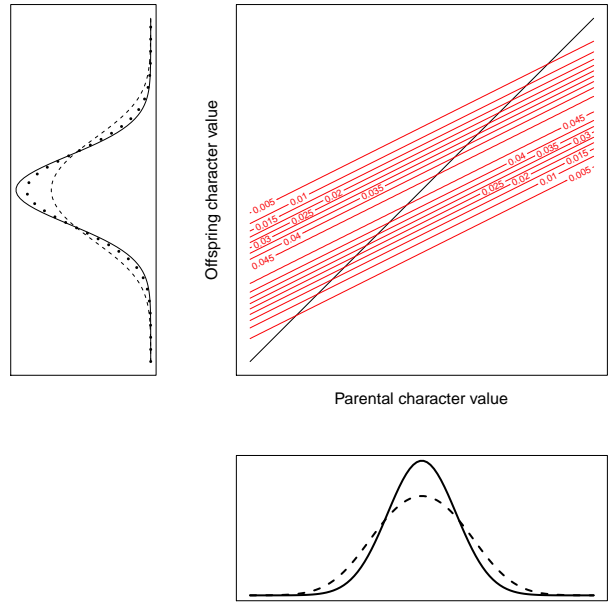}
\caption{\label{fig:fig3} Dynamics of a transmission function.  Start (bottom panel) with a normal (solid line) and non-normal (dashed line) distribution.  Multiply the distributions by the haploid function (black line) or the diploid function (red kernel). The haploid map ensures that the parental and offspring distribution are identical. This Gaussian diploid map generates an offspring distribution identical to the parental distribution for the normal distribution.  However, for the non-normal distribution this is not the case, with the transmission function injecting variation into the offspring distribution (dotted line, left panel).  The initial distribution is shown to aid comparison.}
\end{figure}

\clearpage

\begin{figure}
\centering
\includegraphics[width=9cm]{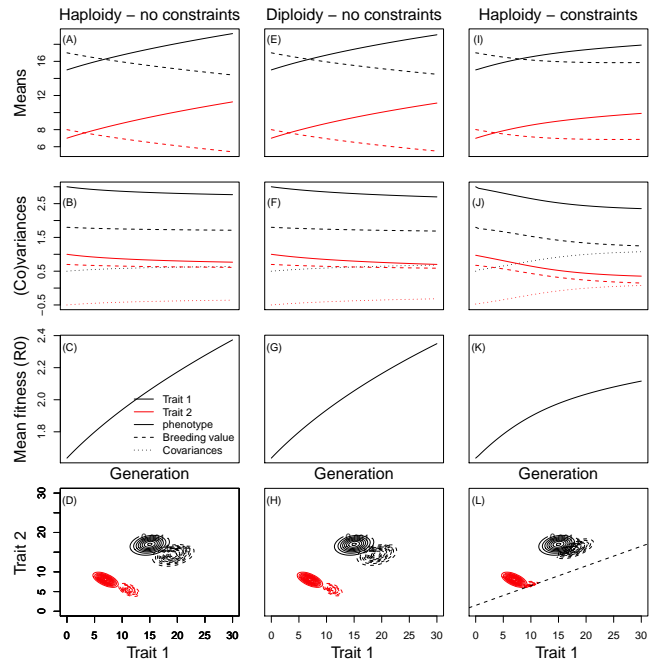}
\caption{\label{fig:fig4} A dynamic version of the breeders equation. Changes in the mean of the phenotype and breeding value distribution (A), (E) and (I), the phenotypic and genetic (co)variances (B), (F) and (J), mean fitness (C), (G) and (K) and representations of the breeding value (red) and phenotype (black) distributions at the start (solid contours) and end (dashed contours) of 60 generation simulations for the haploid model without any genetic constraints (A)-(D), the diploid model without any genetic constraints (E)-(H) and the haploid model with genetic constraints (I)-(L). The legend represents line colours and type for the dynamics of the means and covariances.}
\end{figure}

\clearpage

\begin{figure}
\centering
\includegraphics[scale=0.5]{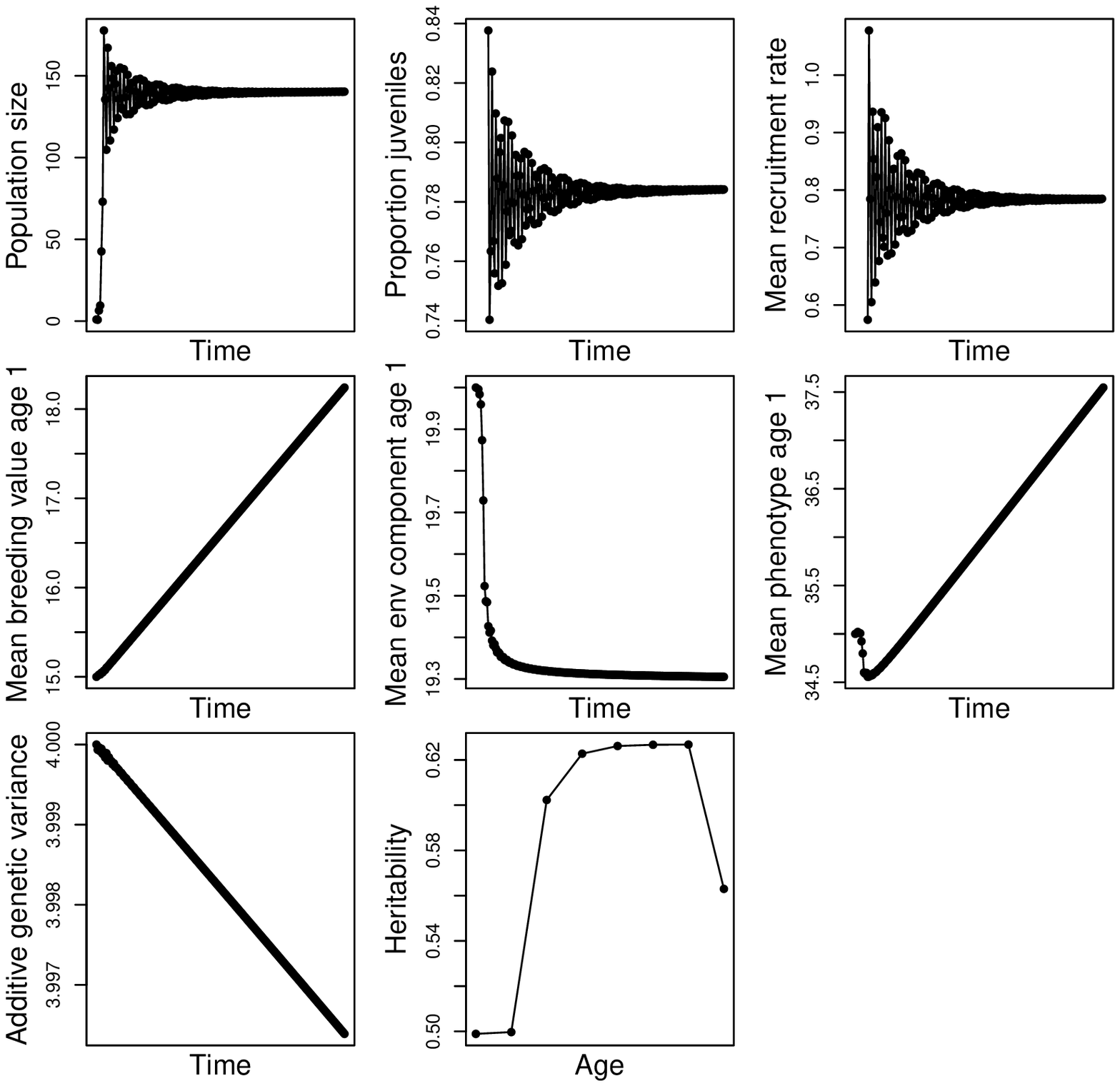}	
\end{figure}

\begin{figure}
\centering
\includegraphics[width=9cm]{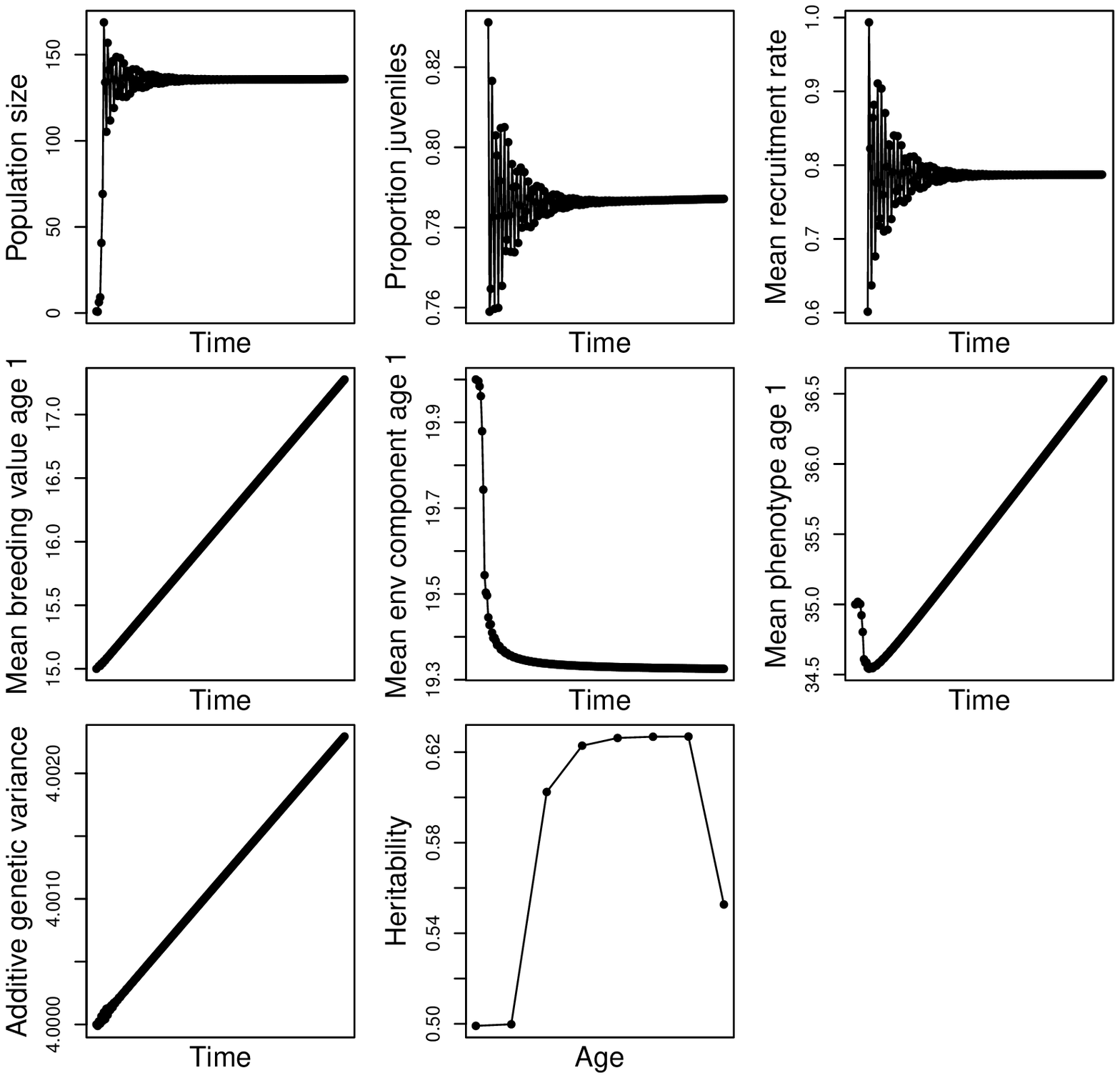}	
\end{figure}

\begin{figure}
\centering
\includegraphics[width=9cm]{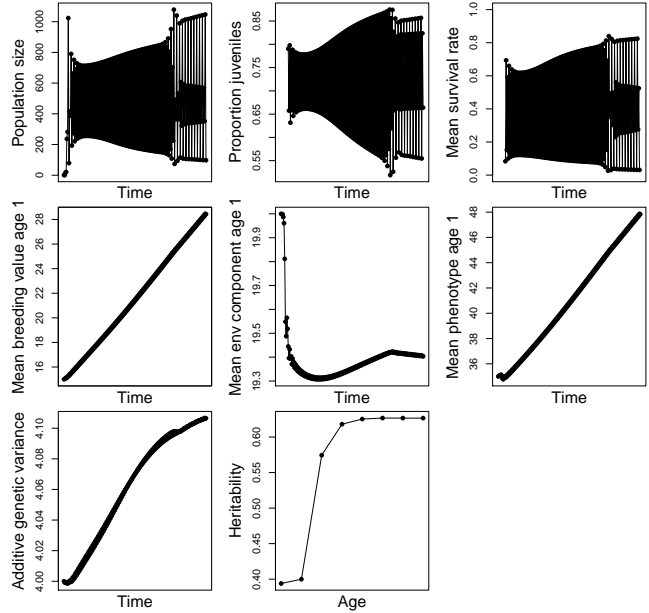}	
\caption{\label{fig:fig5c} Dynamics of the age-structured, density-dependent model parameterised for (a) the baseline case (b) the transition between population dynamic patterns, and (c) with changes in the development functions for the breeding value.  In the top row of panels in each figure we show how population size, population structure (measured as proportion juveniles) and mean survival change through the simulation.  In the middle row of panels we show how the mean breeding value, environmental component of the phenotype and the phenotype evolve.  In the third row of the panel we report trends in the additive genetic and how the heritability changes with age.}
\end{figure}

\end{document}